\documentclass{ifacconf}

\usepackage{graphicx}
\usepackage{natbib}
\usepackage{xcolor}
\usepackage{amsmath,amssymb,amsfonts}
\usepackage{algorithm}
\usepackage{algpseudocode}
\usepackage{soul}
\usepackage{mathtools}

\usepackage{silence}
\usepackage{subcaption}

\newcommand{\Rspace}{\mathbb{R}}

\renewcommand{\d}{\partial}
\renewcommand{\st}{\text{s.t.  }}
\newcommand{\norm}[1]{\left\lVert #1 \right\rVert}
\newcommand{\abs}[1]{\left\lvert #1 \right\rvert}

\newcommand{\X}{\mathcal{X}}
\newcommand{\Y}{\mathcal{Y}}
\newcommand{\Z}{\mathcal{Z}}
\newcommand{\T}{\intercal}

\begin{document}
\begin{frontmatter}

\title{Error Bounds for Compositions of Piecewise Affine Approximations\thanksref{footnoteinfo}} 
\thanks[footnoteinfo]{This work was supported by the Department of Defense through the National Defense Science \& Engineering Graduate (NDSEG) Fellowship Program.}

\author{Jonah J. Glunt,} 
\author{Jacob A. Siefert,} 
\author{Andrew F. Thompson,}
\author{Herschel C. Pangborn}

\address{Department of Mechanical Engineering\\The Pennsylvania State University, University Park, PA 16802 USA\\(e-mails: jglunt@psu.edu; jas7031@psu.edu; thompson@psu.edu; hcpangborn@psu.edu).}

\begin{abstract}
    Nonlinear expressions are often approximated by piecewise affine (PWA) functions to simplify analysis or reduce computational costs. 
To reduce computational complexity, multivariate functions can be represented as compositions of functions with one or two inputs, which can be approximated individually. 
This paper provides efficient methods to generate PWA approximations of nonlinear functions via functional decomposition. 
The key contributions focus on intelligent placement of breakpoints for PWA approximations without requiring optimization, and on bounding the error of PWA compositions as a function of the error tolerance for each component of that composition. The proposed methods are used to systematically construct a PWA approximation for a complicated function, either to within a desired error tolerance or to a given level of complexity. 
\end{abstract}
\begin{keyword}
Hybrid and switched systems modeling, verification of hybrid systems, error quantification
\end{keyword}

\end{frontmatter}

%========================================

\section{Introduction}
\label{sect:introduction}

Piecewise affine (PWA) approximations are used to reduce computational costs in system identification, nonlinear control, and reachability of nonlinear systems~\citep{bemporad_observability_2000,asarin_approximate_2000,asarin2003reachability,roll2003local,Bemporad2005,baier2007approximation,lai2010identification, lai2011data,LatVehicleBorrelli2012}. Additionally, neural networks with PWA activation functions have been trained to approximate nonlinear functions~\citep{wang2008_NNs_and_PWA, guhring2019error}. Every continuous multivariate function can be equivalently expressed as the composition and addition of continuous functions of a single variable \citep{Kolmogorov1957}. The general process of representing a multivariate function using a composition of simpler functions, referred to as \emph{functional decomposition}, has been leveraged extensively~\citep{Dijkstra1961SYA,leyffer2008branch,SZUCS_2012_OptimalPWA,Siefert_Reach_FuncDecomp}. Constructing PWA approximations for the functional decomposition avoids the exponential scaling associated with approximating in higher-dimensional spaces directly, though analyzing the propagation of error through functional compositions of PWA approximations is nontrivial.

Several methods to generate PWA approximations solve nonlinear optimization programs to minimize various forms of approximation error, such as least-squares or maximum norm \citep{dunham1986optimum, magnani2009convex, KOZAK_2011, SZUCS_2012_OptimalPWA, Rebennack2015, Diop2023}. \cite{ihm1991piecewise} present multiple heuristic solutions as an alternative to optimization programs. In~\cite{SZUCS_2012_OptimalPWA}, a class of \emph{multivariable separable functions} are shown to be well-suited for PWA approximation, and the authors provide nonlinear programs to approximate unary or binary function within the composition. Alternatively, neural networks can be trained to approximate a nonlinear function, and then a PWA approximation of the activation functions can be applied \citep{KOZAK_2011, guhring2019error}. Additionally, \cite{IMAI_1986,IMAI_1988} present algorithms using graph theory to reduce the complexity of a PWA approximation, given an error tolerance or a desired number of breakpoints. An overview of most of these methods is given in \cite{Groot2013}.

Error bounds of PWA approximations for various tilings of input spaces are presented in \cite{pottmann2000piecewise, kristensen2016piecewise}. Methods proposed in~\cite{Zavieh2013} analyze the PWA approximation error of convex nonlinear systems. For nonconvex functions, the domain is split at inflection points. There exist methods to bound the error in approximating a nonlinear function with an affine interpolation over a simplex using a Lipschitz constant~\citep{azuma2010lebesgue}, or other classes of functions such as $\mathcal{C}^0$, $\mathcal{C}^1$, and $\mathcal{C}^2$~\citep{Stampfle2000}. 

\subsubsection{Contributions}
This paper first contributes a method to generate PWA approximations of nonlinear systems that intelligently places breakpoints to achieve a desired error tolerance without requiring optimization. It then focuses on quantifying how error propagates through compositions of piecewise affine approximations of several classes of functions. This enables efficient construction of PWA approximations to a given error tolerance and/or complexity, supporting their use for efficient analysis and control of nonlinear systems. 

\section{Notation and Preliminaries}
\label{sect:Prelims}

Matrices are denoted by uppercase letters, e.g., $G\in\Rspace^{n\times m}$, and sets by uppercase calligraphic letters, e.g., $\mathcal{Z}\subset\Rspace^{n}$. 
Vectors and scalars are denoted by lowercase letters. 
The concatenation of two column vectors to a single column vector is denoted by $(g_1,\:g_2)=[g_1^\T\:g_2^\T]^\T$.
A vector inside of single vertical lines  $\abs{g}$ is used to denote an element-wise absolute value, while p-norms are notated with double vertical lines $\norm{g}_p$. 
A function $f(\cdot)$ is said to be $\mathcal{C}^n$ if it is continuous and $n$-times differentiable over a given domain. 
Lagrange notation $f'(\cdot)$ denotes the derivative of a real-valued function of one real variable, with Leibniz notation $\frac{\d f}{\d y}$ used for scalar functions with multivariable inputs or when the variable of differentiation is not clear from context. 
Given a function $f:\Rspace^n\rightarrow\Rspace^m$, the graph of the function $\mathcal{G}_f$ is defined as $\mathcal{G}_f=\{(x,y)\ \vert\ y=f(x)\}$. 
A closed interval $\mathcal{I}\subset\Rspace$ is written in shorthand as $[a,\ b]$ and is defined as $\mathcal{I}=\{x\in\Rspace\ \vert\ a\leq x\leq b\}$.

We adopt the following definition from~\cite{SZUCS_2012_OptimalPWA}: A function $f:\Rspace^n\rightarrow\Rspace$ is \emph{piecewise affine} if and only if
\begin{align}
    \label{eqn:PWA_defn}
    f(x) = \begin{cases*}
        a_1^\T x+b_1 & if $x\in\mathcal{R}_1$ \\
         & \vdots\\
        a_N^\T x + b_N & if $x\in\mathcal{R}_N$
    \end{cases*}\;,
\end{align}
where $\text{int}(\mathcal{R}_i)\cap\text{int}(\mathcal{R}_j)=\varnothing\ \forall i\neq j$ and domain $\mathcal{D}=\bigcup_i \mathcal{R}_i$, $i\in\{1,\cdots,N\}$. Furthermore, we assume that each $\mathcal{R}_i$ is a convex polytope and refer to the vertices of those polytopes as \emph{breakpoints}.

PWA approximations are closely related to special ordered set (SOS) approximations~\citep{beale1970_SOS}. Specifically, SOS approximations are a special case of PWA approximations where the domain is separated into a collection of simplices and each breakpoint lies on the graph of the function; i.e., if $\mathcal{S}$ is an SOS approximation of function $f$ and has a vertex at $(x,y)\in\mathcal{S}$, then $y=f(x)$. The PWA approximations constructed in this paper need not be SOS approximations except where explicitly stated.

All numerical examples in this paper were implemented in MATLAB on a desktop computer with a 3.0 GHz Intel i7 processor and 16 GB of RAM, using \texttt{fmincon} and \texttt{GUROBI}~\citep{gurobi_optimization_gurobi_2021} when required.

\section{PWA Approximations with Specified Error Tolerance}
\label{sect:PWAApprox}

This section provides methods of constructing PWA approximations of continuous nonlinear functions ($\mathcal{C}^0$) that map $\Rspace\rightarrow\Rspace$ with a user-defined error tolerance and interval domain. When constructing PWA approximations, the resulting memory complexity, defined by the number of breakpoints, and the computational cost of constructing the PWA approximation must be considered. Two methods are proposed. Similar to techniques in the literature, Method 1 produces PWA approximations with minimal resulting memory complexity but requires nonlinear optimization, which could be computationally expensive to apply \citep{dunham1986optimum, magnani2009convex, KOZAK_2011, SZUCS_2012_OptimalPWA, Rebennack2015, Diop2023}. Alternatively, Method 2 provides better computational efficiency by not requiring this optimization, but may require more breakpoints for the same tolerance and involves more restrictive assumptions.

\subsection{Method 1: Optimal Placement of Breakpoints}
Method 1, given by Algorithm~\ref{alg:SOSstroll}, uses bisection to generate each successive breakpoint such that the specified error tolerance is satisfied. \textbf{Lines 1-2} initialize the index $k$, first breakpoint $x_1$, and assign $e$ to the maximum error associated with a domain spanning $[x_1,\Bar{x}]$. The evaluation of the maximum error using the function $\texttt{evalErr}(f(\cdot),[x_1,m])$ is defined in general by
\begin{align}
    \label{eqn:evalErrdef}
    \texttt{evalErr}(f(\cdot),[x_k,m]) = \max_{x\in[x_k,m]} |\Bar{f}(x)-f(x)|\:,
\end{align}%
where $\Bar{f}(x)$ is given by the PWA (secant line) approximation on the domain $[x_k,m]$ as
\begin{align}
    \bar{f}(x) = f(x_k) + \frac{f(m)-f(x_k)}{m-x_k}(x-x_k)\:.
\end{align}%
\textbf{Line 3} checks if $\Bar{x}$ is the last breakpoint. \textbf{Line 4} initializes upper, lower, and cut values for a bisection method. \textbf{Line 5} checks if the bisection method has converged. \textbf{Lines 6-11} evaluate the maximum error over the region $[{x_k,m}]$ and reassign bisection upper, lower, and cut values appropriately. \textbf{Line 12} assigns the next breakpoint  $x_{k+1}$ to the lower limit $\ell$ (for tolerance satisfaction), and updates the error for the domain spanning $[x_{k+1},\Bar{x}]$. \textbf{Line 13} increments $k$ and assigns $x_{k+1}$ to the upper domain limit $\bar{x}$.

In general, evaluating \eqref{eqn:evalErrdef} requires solving a nonlinear program. Additional properties of the function $f$, such as whether $f(x)$ is differentiable, convex, or piecewise-defined with differentiable pieces, can significantly reduce the complexity of solving \eqref{eqn:evalErrdef}. 

\begin{algorithm}
Result: Vector of breakpoints $x$
\begin{algorithmic}[1]
    \State $k\leftarrow1$, $x_1 \leftarrow \underline{x}$
    \State $e \leftarrow \texttt{evalErr}(f(\cdot),[x_1,\bar{x}])$
    \While{$e>\tau$}
        \State $\ell\leftarrow x_k$, $u \leftarrow \bar{x}$, $m \leftarrow (\ell+u)/2$
    \While{$u-\ell>\delta_x$ OR $e>\tau$}
        \State $e \leftarrow \texttt{evalErr}(f(\cdot),[x_k,m])$
        \If{$e<\tau$}
            $\ell \leftarrow m$
        \ElsIf{$e>\tau$}
            $u \leftarrow m$
        \Else{
            $\ell \leftarrow m$, $u \leftarrow m$}
        \EndIf
        \State $m \leftarrow (\ell+u)/2$
    \EndWhile
    \State $x_{k+1} \leftarrow \ell$, $e \leftarrow \texttt{evalErr}(f(\cdot),[x_{k+1},\bar{x}])$
    \State $k \leftarrow k+1$, $x_{k+1} \leftarrow \bar{x}$ \nonumber
    \EndWhile
\end{algorithmic}
     \caption{Generate PWA approximation. Given: Continuous unary scalar-valued function $f(\cdot)$, output error tolerance $\tau$, breakpoint tolerance $\delta_x$, input domain $[\underline{x},\ \bar{x}]$.}
     \label{alg:SOSstroll}
\end{algorithm}

\begin{exmp}
    Consider the function $y=\sin (x)$ over the domain $x\in[0,\ 2\pi]$. Figure~\ref{fig:M2_bisection} shows how Algorithm~\ref{alg:SOSstroll} performs a bisection method to generate a PWA approximation with tolerance $\tau=0.3$.
\end{exmp}
\newcommand\widI{2.75in}
\begin{figure}
    \centering
    \begin{subfigure}[b]{\widI}
        \includegraphics[width=\textwidth]{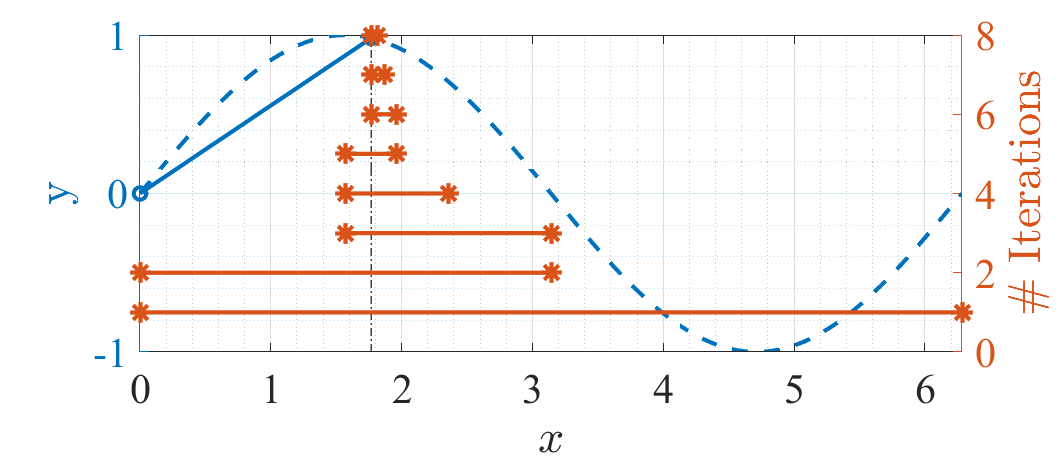}
        \caption{}
        \label{$x_2$}
    \end{subfigure}\\
    \begin{subfigure}[b]{\widI}
        \includegraphics[width=\textwidth]{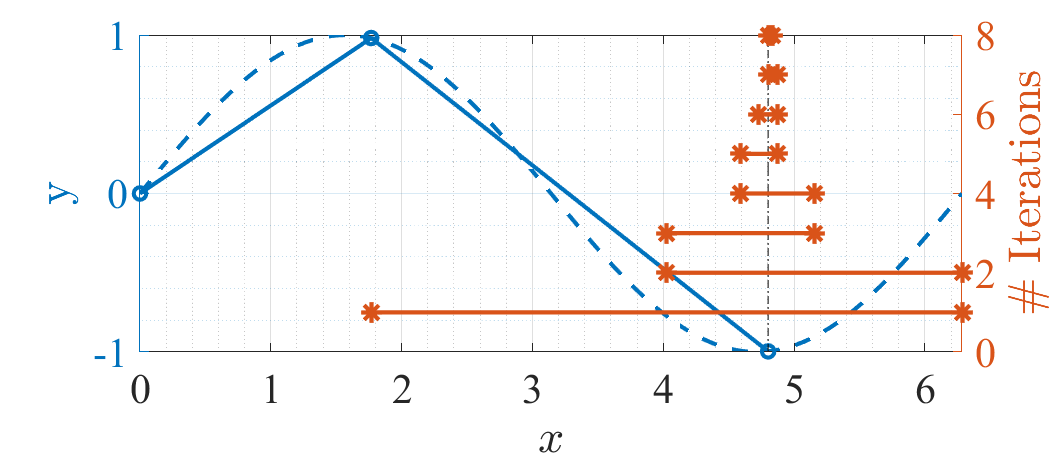}
        \caption{}
        \label{$x_3$}
    \end{subfigure}\\
    \begin{subfigure}[b]{\widI}
        \includegraphics[width=\textwidth]{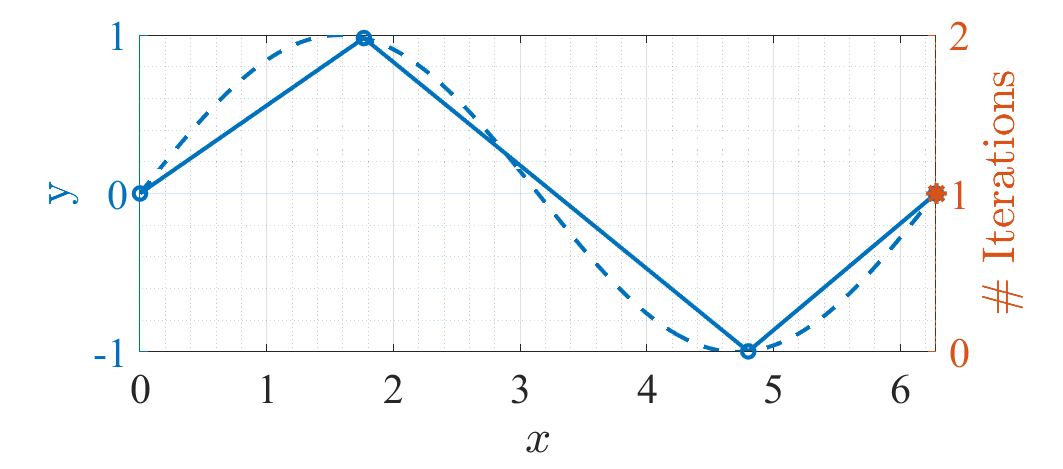}
        \caption{}
        \label{$x_4$}
    \end{subfigure}\\
    \caption{Construction of PWA approximation of $y=\sin (x)$  using Method~1. (a) Bisection to determine $x_{2}$. (b) Bisection to determine $x_3$. (c) The algorithm concludes with a breakpoint at the upper limit of the domain $\Bar{x}=2\pi$, for which the  the error tolerance is satisfied without requiring iteration.}
    \label{fig:M2_bisection}
\end{figure}

\subsection{Method 2: Algebraic Calculation of Breakpoints}
Algorithm~\ref{alg:SOSsprint} assumes that the function $f(\cdot)$ is $\mathcal{C}^3$ over the domain $[\underline{x},\ \bar{x}]$, and a bound on the magnitude of the third derivative over the domain is given by 
\begin{align}
\label{eqn:d3def}
    d_3\geq \max_{x\in[\underline{x},\bar{x}]}|f'''(x)|\:. 
\end{align}%
Using these assumptions, Algorithm~\ref{alg:SOSsprint} leverages a \emph{closed-form bound} on the error presented in Theorem~\ref{thm:C2thriceDiff_errBound}.
\begin{thm}
    \label{thm:C2thriceDiff_errBound}
    Given a unary scalar-valued function $f(\cdot)$ that is $\mathcal{C}^3$ over $x\in[\underline{x},\Bar{x}]$, and a bound on the magnitude of the third derivative over the domain $d_3$  \eqref{eqn:d3def}, then
    \begin{align}
    \label{eqn:SOSerrBound_Cubic}
        \max_{x\in[\underline{x},\Bar{x}]} |\Bar{f}(x)-f(x)| \leq \frac{d_3}{8}(\Bar{x}-\underline{x})^3 + \frac{d_2}{8}(\Bar{x}-\underline{x})^2\:,
    \end{align}%
    where $d_2=|f''(\underline{x})|$.
\end{thm}
\begin{pf}
    The proof begins with the result in Theorem 4.1 of \cite{Stampfle2000}, modified to match notation herein and specified to $f:\Rspace\rightarrow\Rspace$,
    \begin{align}
    \label{eqn:C2bound_Stampfle}
         \max_{x\in[\underline{x},\Bar{x}]} |\Bar{f}(x)-f(x)| \leq \frac{1}{8}\max_{x\in[\underline{x},\bar{x}]} |f''(x)|(\Bar{x}-\underline{x})^2\:.
    \end{align}%
    By the fundamental theorem of calculus and repeated use of the triangle inequality,
        \begin{align}
            \nonumber
            f''(\Bar{x})&=f''(\underline{x}) + \int_{\underline{x}}^{\Bar{x}} f'''(t) \,dt \;,\\
            %%%
            \nonumber
            \abs{f''(\Bar{x})} &= \abs{ f''(\underline{x}) + \int_{\underline{x}}^{\Bar{x}} f'''(t) \,dt } \;,\\
            %%%
            \nonumber
            &\leq \abs{f''(\underline{x})} + \int_{\underline{x}}^{\Bar{x}} \abs{f'''(t)} \,dt \;,\\
            %%%
            \label{eqn:max_fppx}
            &\leq \abs{f''(\underline{x})} + \max_{t\in[\underline{x},\Bar{x}]}\abs{f'''(t)}(\Bar{x}-\underline{x})\;.
        \end{align}%
    Replacing the maximum magnitude of the second derivative in \eqref{eqn:C2bound_Stampfle} with an upper bound given by \eqref{eqn:max_fppx} yields
    \eqref{eqn:SOSerrBound_Cubic}. 
    \qed
\end{pf}

Algorithm~\ref{alg:SOSsprint} is summarized as follows.
\textbf{Line~1} initializes index $k$, first breakpoint $x_1$, and magnitude of the second derivative at the current breakpoint.
\textbf{Line~2} assigns $e$ to the error bound over $[x_1,\Bar{x}]$.
\textbf{Line~3} checks if $\bar{x}$ is the last breakpoint. If not, 
\textbf{lines~4-5} assign the next breakpoint $x_{k+1}$ to the largest real solution of 
\begin{equation}
    \frac{d_3}{8}(x_{k+1}-x_{k})^3 + \frac{d_2}{8}(x_{k+1}-x_{k})^2 = \tau\:,
\end{equation}
which falls in the domain $[\underline{x},\Bar{x}]$.
\textbf{Lines~6-9} update $e$ and $d_2$ associated with the last breakpoint $x_{k+1}$ and increment $k$.
\textbf{Line~10} assigns $\Bar{x}$ as the final breakpoint.

\begin{rem}
    The variation in breakpoint separation is obtained by updating $d_2$, e.g., when $d_2$ is small, the next $x_{k+1}$ can be further away. This offers improvement in complexity over uniform breakpoint separation.
\end{rem}

\begin{algorithm}
Result: Vector of breakpoints $x$
\begin{algorithmic}[1]
    \State $k\leftarrow1$, $x_1 \leftarrow \underline{x}$, $d_2 \leftarrow |f''(x_1)|$
    \State $e \leftarrow \frac{d_3}{8}(\bar{x}-x_1)^3 + \frac{d_2}{8}(\bar{x}-x_1)^2$
    \While{$e>\tau$}
        \State $\{s_1,s_2,s_3\}\leftarrow \texttt{roots}(\frac{d_3}{8}(x-x_{k})^3 + \frac{d_2}{8}(x-x_{k})^2 - \tau)$
        \State $x_{k+1} = \max(s)$ s.t. $s\in[\underline{x},\bar{x}]\cap\{s_1,s_2,s_3\}\cap\Rspace $
        \State $d_2 \leftarrow |f''(x_{k+1})|$
        \State $e \leftarrow \frac{d_3}{8}(\bar{x}-x_{k+1})^3 + \frac{d_2}{8}(\bar{x}-x_{k+1})^2$
        \State $k \leftarrow k+1$
    \EndWhile
    \State $x_{k+1} \leftarrow \bar{x}$
\end{algorithmic}
    \caption{Generate PWA approximation. Given: unary scalar-valued function $f(\cdot)$ that is $\mathcal{C}^3$ over $x\in[\underline{x},\ \bar{x}]$, output error tolerance $\tau$, and a bound on the magnitude of the third derivative $d_3~\geq~\max_{x\in[\underline{x},\Bar{x})]} |f'''(x)|$.}
     \label{alg:SOSsprint}
\end{algorithm}

Figure \ref{fig:M1vM2_SOS_TimeAndBreak} compares the construction of PWA approximations using Method 1 and Method 2 in terms of computation time and the number of breakpoints for the functions and domains given in Table~\ref{tab:M1vM2_SOS_FuncAndDomain}. Method 1 is slower to compute but produces approximations with fewer breakpoints.
\begin{table}[ht!]
    \centering
    \caption{Functions and domains used for comparison of Method 1 and Method 2 in Figure~\ref{fig:M1vM2_SOS_TimeAndBreak}.}
        \begin{tabular}{c||c|c|c|c}
        Function & $\sin(x)$ & $x^2$ & $x^3$ & $1/x$ \\ \hline
        Domain & $x\in[0, 2\pi]$ & $x\in[-5, 5]$ & $x\in[-5, 5]$ & $x\in[1, 10]$
    \end{tabular}\label{tab:M1vM2_SOS_FuncAndDomain}
\end{table}

\newcommand\widII{2.5in}
\begin{figure}[ht!]
    \centering
    \includegraphics[width=3in]{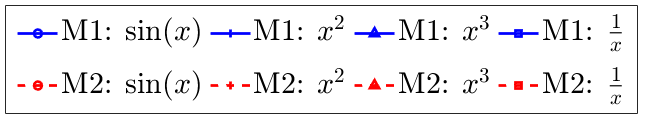}  
    \begin{subfigure}[b]{\widII}
        \centering
        \includegraphics[width=\textwidth]{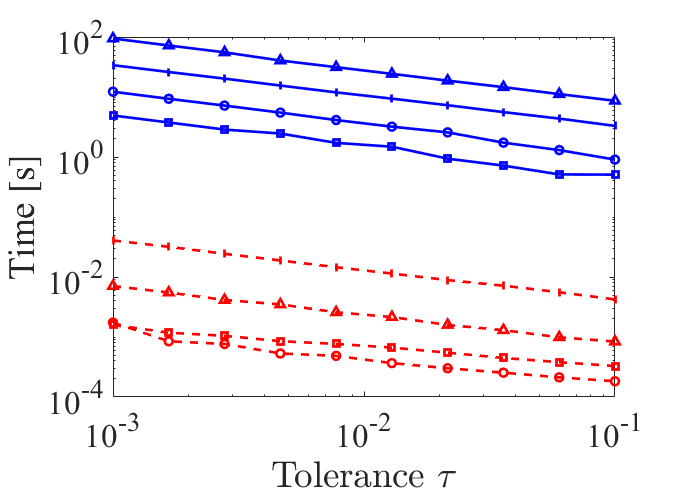}
        \caption{}
    \end{subfigure}
    \begin{subfigure}[b]{\widII}
        \centering
        \includegraphics[width=\textwidth]{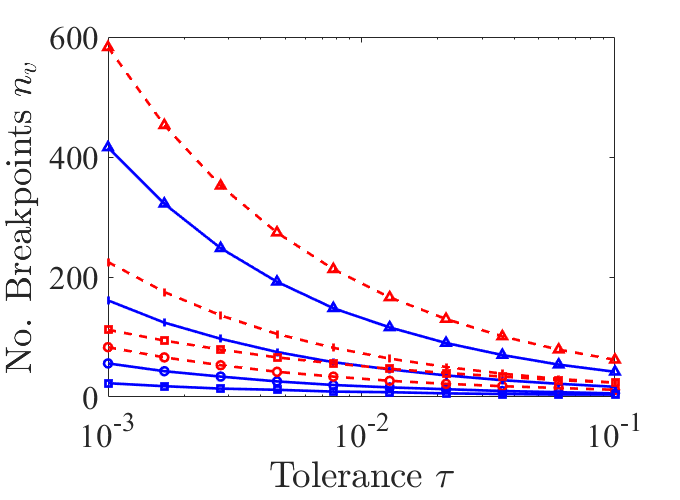}
        \caption{}
    \end{subfigure}
    \caption{ Comparison of Method~1 and Method~2 for functions and domains given in Table~\ref{tab:M1vM2_SOS_FuncAndDomain}. Figure~\ref{fig:M1vM2_SOS_TimeAndBreak}(a) and Figure~\ref{fig:M1vM2_SOS_TimeAndBreak}(b) show the number of breakpoints and computation time as a function of tolerance, respectively. Method 1 is significantly slower to compute than Method 2, but constructs an approximation to a specified error tolerance using fewer breakpoints.}
    \label{fig:M1vM2_SOS_TimeAndBreak}
\end{figure}

\begin{figure}
\centering
    \includegraphics[width=3in]{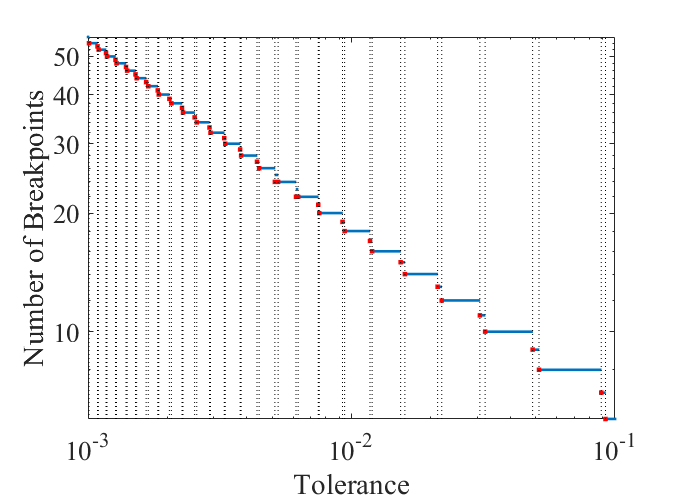}
    \caption{Number of PWA approximation breakpoints required to meet a specified tolerance for the function $\sin(x),\ x\in[0,2\pi]$ using Method~1. Dashed lines indicate where the number of breakpoints ``jumps''.}
    \label{fig:NbreakJumpTol}
\end{figure}

\begin{rem}
    \label{rem:breakpoint_jump}
    Because the number of breakpoints is discrete, as the tolerance is varied the number of breakpoints required to achieve the tolerance will ``jump'' at critical tolerance values. This is demonstrated in Figure~\ref{fig:NbreakJumpTol}, where PWA approximations of $\sin(x)$, $x\in[0,2\pi]$ were obtained via Method~1 by logarithmically sampling the tolerance space 500 times.
\end{rem}

\section{Error Propagation in Compositions of PWA Approximations}
\label{sect:ErrorProp}

When approximating complicated functions, it can be beneficial to decompose the function into unary and binary functions, build a PWA approximation for each, and then compose these to approximate the original function~\citep{Siefert_Reach_FuncDecomp}. However, in addition to generating this approximations itself, it is also desirable to quantify how approximation error propagates through the composition. We now turn our attention to functional compositions of the form $f(g(x)),\ x\in\X\subseteq\Rspace^{n_x}$, which can be applied iteratively to achieve an arbitrary number of compositions. Specifically, given the PWA approximations $\bar{f}(\cdot)\approx f(\cdot)$ and $\bar{g}(\cdot)\approx g(\cdot)$, we seek to bound the compositional error
\begin{align}
    \label{eqn:approx_error}
     \max_{x\in\X} \abs{\bar{f}(\bar{g}(x))-f(g(x))} \leq \varepsilon_{f\circ g}\;.
\end{align}
Existing methods in the literature can be used to bound the error $\abs{\bar{f}(\cdot)-f(\cdot)}$ in terms of the gradient of $\bar{f}$, however this dependence on the approximation makes it difficult to analyze the sensitivity of the error to the tolerance of the approximation. Furthermore, to the authors' knowledge, previous work has not addressed how the error tolerances of $\bar{f}$ and $\bar{g}$ propagate through their composition $\bar{f}(\bar{g}(\cdot))$. 

We now present four theoretical results, the culmination of which yields an error bound for~\eqref{eqn:approx_error} that only depends on the derivatives of $f$ and $g$ as well as the tolerance of their individual approximations. 
Theorem~\ref{thm-C1err_bound} is similar to results in~\cite{azuma2010lebesgue} in that it bounds the approximation error in terms of the gradient of the approximation, however unlike this literature it focuses on quantifying how approximation error propagates through composition. 
Corollary~\ref{cor:PWA_err_bound} extends this result to a PWA approximation. Corollaries~\ref{cor_unary_err_bound} and~\ref{cor:MVT_err_bound_PWA} then yield an error bound that is only a function of the tolerances of the individual approximations. 

\begin{thm}
    \label{thm-C1err_bound}
    Consider $x\in\X\subseteq\Rspace^{n_x}$, $y\in\Y\subseteq\Rspace^{n_y}$, $z\in\Z\subseteq\Rspace$, function $g(\cdot): \X\rightarrow\Y$, function $f(\cdot): \Y\rightarrow\Z$, and the functional composition $f(g(x))$. Assume the following are given: a) an approximation $\Bar{g}(x) \approx g(x)$ with a vector of error bounds $\varepsilon_{g}$ such that 
    \begin{align}
        \label{thm-C1ErrProp-eqn-gTol}
        \max_{x\in\X}\ \abs{\Bar{g}_i(x)-g_i(x)}\ \leq \varepsilon_{g,i}\ \forall\ i\in\{1,...,n_y\}\:,
    \end{align}%
    where the subscript $i$ refers to the $i$-th element of the vectors $\bar{g},\ g,\ $and $ \varepsilon_g$, 
    b) an affine approximation
    \begin{align}
        \label{eqn:C1errBd_fbar_affine}
        \Bar{f}(y) = a^\T y + b \approx f(y)
    \end{align}
    over the domain $y\in\Y$ with error bound $\tau_{f}\geq0$ such that
    \begin{align}
        \label{eqn-C1errBdF}
        \max_{y\in\Y}\ \abs{\Bar{f}(y)-f(y)}\ \leq \tau_{f}\:,
    \end{align}%
    and c) a vector of bounds $d_{\Bar{f}}$ such that,
    \begin{align}
        \label{eqn-C1bdFbar}
        \abs{\frac{\d{\Bar{f}}}{\d y}}=\abs{a}\leq d_{\Bar{f}}
    \end{align}
    An upper bound on the error magnitude of the approximated composition
    \begin{align}
        \max_{x\in\X} \abs{\Bar{f}(\Bar{g}(x))-f(g(x))}\ \leq\  \varepsilon_{f {\tiny \circ} g}\:,
    \end{align}
    is given by
    \begin{align}
        \label{thm-C1EP-eqn-ID}
        \varepsilon_{f {\tiny \circ} g} = \tau_f + 
        d_{\bar{f}}^\T \varepsilon_g\;.
    \end{align}
\end{thm}
\begin{pf}
    By \eqref{thm-C1ErrProp-eqn-gTol}, $\forall x\in\X$, $ \exists\:\delta$ such that $\abs{\delta}\leq \varepsilon_g$ and
    \begin{align}
        \label{thm-C1-gbarsub}
        \bar{g}(x) = g(x) + \delta\:.
    \end{align}%
    By construction,
    \begin{align}
        \label{thm-C1ErrProp-eqn1}
        \big\lvert\bar{f} ( \bar{g} (x) ) - f(g(x))\big\rvert &= \abs{a^\T \bar{g}(x) + b - f(g(x))}\:,\\
        %%% 
        \label{thm-C1ErrProp-eqn2}
        &= \abs{a^\T (g(x) + \delta )+ b - f(g(x))},\\
        %%% 
        \label{thm-C1ErrProp-eqn3}
        &\leq \abs{\bar{f}(g(x))-f(g(x))} + \abs{a^\T  \delta}\:,\\
        %%% 
        \label{thm-C1ErrProp-eqn4}
        &\leq \tau_f + \abs{a}^\T  \abs{\delta}\:,\\
        %%% 
        \label{thm-C1ErrProp-eqn6}
        &\leq \tau_f + d_{\Bar{f}}^\T  \varepsilon_g\:.
    \end{align}%
    The first step~\eqref{thm-C1ErrProp-eqn1} is achieved by substitution of~\eqref{eqn:C1errBd_fbar_affine}, and the second step by substitution of~\eqref{thm-C1-gbarsub}. The triangle inequality and substitution of~\eqref{thm-C1ErrProp-eqn-gTol} and~\eqref{eqn-C1errBdF} yields~\eqref{thm-C1ErrProp-eqn6}.
    \qed
\end{pf}
\

\begin{exmp}
\label{ex:ErrProp_affOnly}
Consider the function
\begin{align}
    \label{ex-ErrPropAffOnly-eqn-y}
    f(x) = \left(\sin\left(\frac{1}{x}\right)\right)^2\:,
\end{align}%
on the domain $x\in[1,\ 3]$. 
The functional decomposition variables $w_i$ and their domains are given by
\begin{subequations}
\label{eqn:ex2_decomp}
\begin{align}
    %%%
    w_1 &= x\:, &&w_1 \in[1,\ 3]\;,\\
    %%%
    w_2 &= \frac{1}{w_1}\:, &&w_2 \in \left[\frac{1}{3},\ 1\right]\;,\\
    \label{eqn:ExC1compBounds}
    w_3 &= \sin(w_2)\:, &&w_3\in \left[\sin\left(\frac{1}{3}\right),\ \sin(1)\right]\;,\\
    %%%
    w_4 &= w_3^2\:, &&w_4\in \left[\sin^2\left(\frac{1}{3}\right),\ \sin^2(1)\right]\;.
\end{align}%
\end{subequations}
The domains in~\eqref{eqn:ex2_decomp} are calculated using standard interval arithmetic. The PWA approximation of $w_2(w_1)$ is constructed using Method~1 from Section~\ref{sect:PWAApprox}. Using the affine approximations and tolerances of $w_3(w_2)$ and $w_4(w_3)$ given in Table~\ref{tab:ExC1CompositionBounds}, Theorem~\ref{thm-C1err_bound} can be used to bound the composed error in approximating $w_4(w_1)$. Figure~\ref{fig:ErrProp_affOnly} graphs these functional compositions, their approximations, and their error bounds. As can be seen, using the individual tolerances to calculate $\varepsilon_4$ produces an error bound that contains the approximation and original function. These error bounds are quite conservative, as Theorem~\ref{thm-C1err_bound} is restricted to using only affine approximations in each successive composition.

\begin{table}
    \centering
        \caption{Affine approximations, tolerances, and error bounds for Example~\ref{ex:ErrProp_affOnly}.}
    \begin{tabular}{c|c|c|c}
        Decomposition & Approximation & $\tau_f$ & $\varepsilon_i$ \\ \hline\hline
        $w_2$ & n/a & 0.05 & 0.05 \\ \hline
        $w_3$ & $0.771 w_2 + 0.0701$ & $0.0342$ & $0.0728$ \\ \hline
        $w_4$ & $1.170 w_3 - 0.2753$ & $0.0661$ & $0.1512$
    \end{tabular}
    \label{tab:ExC1CompositionBounds}
\end{table}

\newcommand\widIV{3.3in}
\begin{figure}
    \centering
    \begin{subfigure}[b]{\widIV}
        \includegraphics[width=\textwidth]{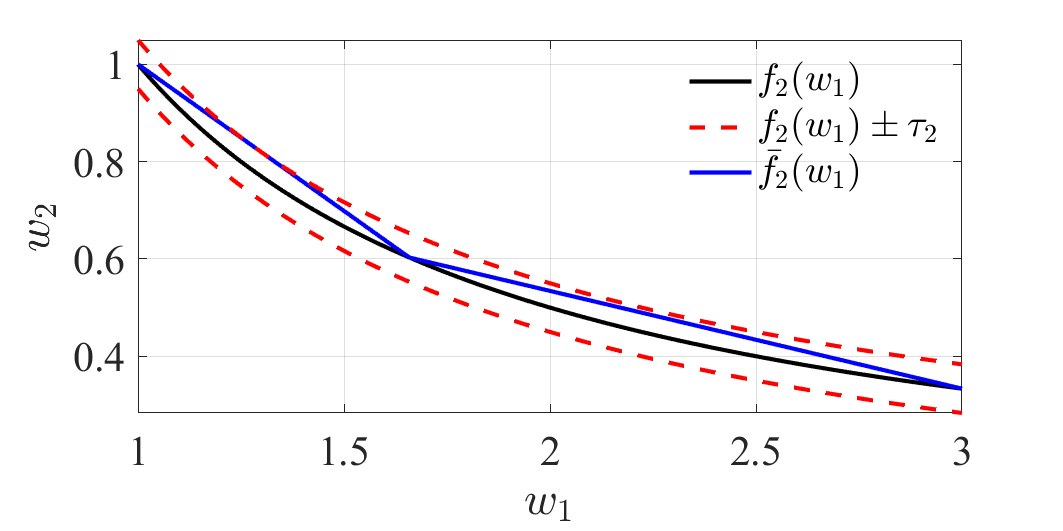}
        \caption{}
    \end{subfigure}\\
    \begin{subfigure}[b]{\widIV}
        \includegraphics[width=\textwidth]{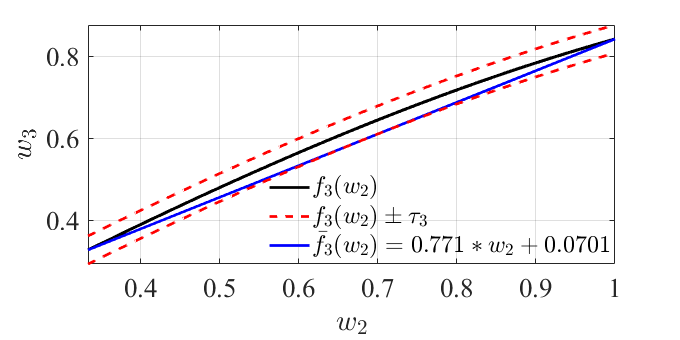}
        \caption{}
    \end{subfigure}\\
    \begin{subfigure}[b]{\widIV}
        \includegraphics[width=\textwidth]{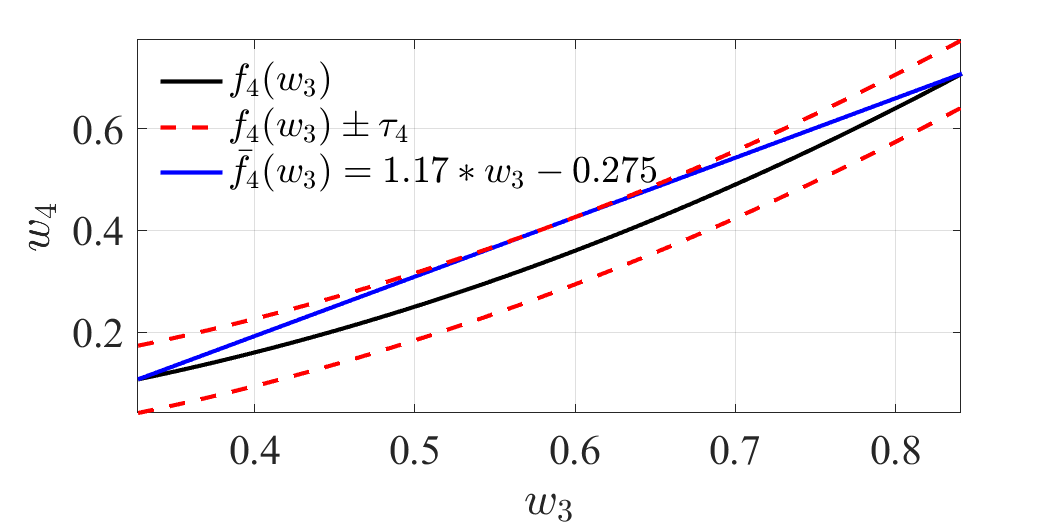}
        \caption{}
    \end{subfigure}\\
    \begin{subfigure}[b]{\widIV}
        \includegraphics[width=\textwidth]{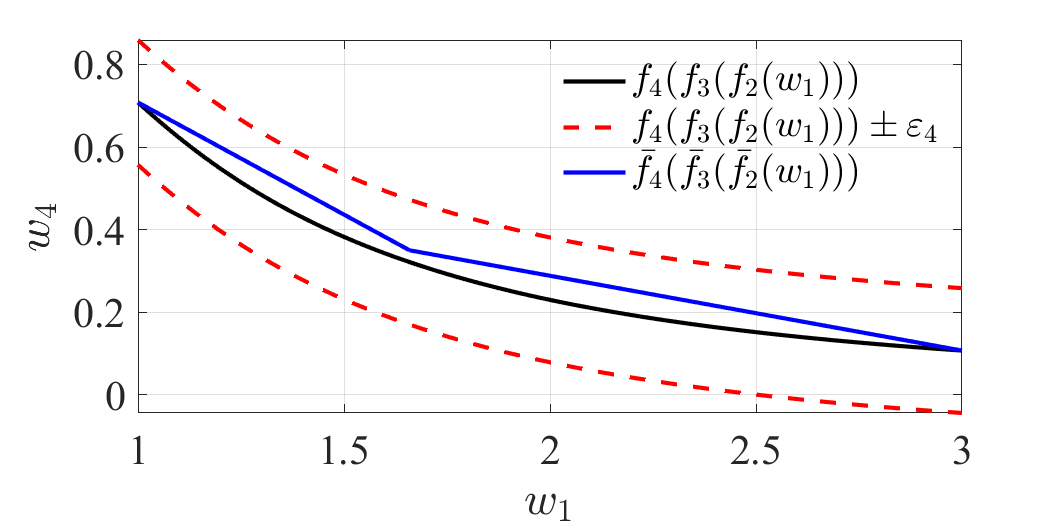}
        \caption{}
    \end{subfigure}
    \caption{Graphs of the decomposed functions and their PWA approximation error bounds from Example~\ref{ex:ErrProp_affOnly}.}
    \label{fig:ErrProp_affOnly}
\end{figure}

\end{exmp}

Theorem~\ref{thm-C1err_bound} bounds the error of composing any approximation $\bar{g}$ into an affine approximation $\bar{f}$. Corollary~\ref{cor:PWA_err_bound} extends this result from an affine approximation of $f$ to a PWA approximation, and Corollary~\ref{cor_unary_err_bound} considers the unary case. Although it may seem limiting to require a unary function for Corollary~\ref{cor_unary_err_bound}, functional decomposition can be used to express any continuous function as the composition and addition of continuous unary functions~\citep{Kolmogorov1957}.
\begin{cor}
    \label{cor:PWA_err_bound}
    Consider the assumptions of~Theorem~\ref{thm-C1err_bound}, however with $\bar{f}(\cdot)$ PWA instead of affine, consisting of $N$ regions with each affine approximation $\bar{f}_j(y)=a_j^\T y + b_j\ \forall\ j\in\{1,...,N\}$, and $\tau_{{f}}$ and $d_{\Bar{f}}$ such that
     \begin{align}
        \label{eqn:cor_PWA_assump1}
        \max_{y\in\Y}\ \abs{\Bar{f}(y)-f(y)}\ &\leq \tau_{f}\:,\text{ and}\\
        %%%
        \label{eqn:cor_PWA_assump2}
        \max_{j\in\{1,...,N\}} \abs{a_j}\ &\leq d_{\Bar{f}}\;.
    \end{align}%
   An upper bound on the error tolerance of the approximated composition
    \begin{align}
        \max_{x\in\X}\ \abs{\Bar{f}(\Bar{g}(x))-f(g(x))}\ \leq \varepsilon_{f {\tiny \circ} g}\:,
    \end{align}
    is given by
    \begin{align}
        \label{cor-C1EP-eqn-ID}
        \varepsilon_{f {\tiny \circ} g} = \tau_f +  d_{\Bar{f}}^\T \varepsilon_{g} \:.
    \end{align}
\end{cor}
\begin{pf}
    $\tau_f$ and $d_{\Bar{f}}$ are valid bounds for all regions, so the proof repeats that of Theorem~\ref{thm-C1err_bound} for each region, and taking the maximum over the regions maintains inequality.
    \qed
\end{pf}

\begin{cor}
    \label{cor_unary_err_bound}
    Consider the assumptions of Theorem~\ref{thm-C1err_bound}, however with $f(\cdot)$ unary, i.e., $n_y=1$. 
    Assume $\exists\;y_1,y_2\in\Y$ with $y_1<y_2$ where $\bar{f}(y_1)=f(y_1)$, such that $\bar{f}(y_2)=f(y_2)$, i.e., $\bar{f}$ is a secant line of $f$. Then, an upper bound on the error magnitude of the approximated composition
    \begin{align}
        \max_{x\in\X}\ \abs{\bar{f}(\bar{g}(x))-f(g(x))}\ \leq\varepsilon_{f\circ g}
    \end{align}
    is given by
    \begin{align}
        \label{eqn:cor_unary_bound}
        \varepsilon_{f\circ g} = \tau_f+\max_{y\in\Y}\ \abs{f'(y)}\varepsilon_g\:.
    \end{align}
\end{cor}
\begin{pf}
    Because this is a special case of Theorem~\ref{thm-C1err_bound}, it still holds that
    \begin{align}
        \label{eqn:cor_unary_eq1}
        \abs{\bar{f}(\bar{g}(x))-f(g(x))}\leq\tau_f+\abs{a}\varepsilon_g\:.
    \end{align}
    By the mean value theorem, $\exists\;y^* $ such that $ y_1<y^*<y_2$ and
    \begin{align}
        \label{egn:cor_unary_eq2_MVT}
        f'(y^*)=\frac{f(y_2)-f(y_1)}{y_2-y_1}=a\:.
    \end{align}
    Substituting into~\eqref{eqn:cor_unary_eq1} yields
    \begin{align}
        \abs{\bar{f}(\bar{g}(x))-f(g(x))}&\leq\tau_f+\abs{f'(y^*)}\varepsilon_g\:\\
        %%%
        &\leq \tau_f + \max_{y\in\Y}\ \abs{f'(y)}\varepsilon_g\:. \qed
    \end{align} 
\end{pf}

Equation~\eqref{eqn:cor_unary_bound} gives an error bound that is not dependent on the gradient of the approximation but only on the original function and the error of input approximations. Corollaries~\ref{cor:PWA_err_bound}~and~\ref{cor_unary_err_bound} combined yield an error bound for composed PWA approximations of unary functions.

\begin{cor}
    \label{cor:MVT_err_bound_PWA}
    Consider the assumptions of Theorem~\ref{thm-C1err_bound}, however with $f(\cdot)$ unary and $\bar{f}(\cdot)$ PWA instead of affine with the following properties: 
    a) $\bar{f}$ consists of $N$ regions where the $j$-th region has interval domain $\Y_j$ with associated affine approximation denoted $\bar{f}_j(y)=a_j y + b_j\;\forall\;j\in\{1,...,N\}$, 
    b)  $\tau_f$ and $d_{\bar{f}}$ such that~\eqref{eqn:cor_PWA_assump1}~and~\eqref{eqn:cor_PWA_assump2} hold, and 
    c) $\forall\;j\in\{1,...,N\}\;\exists\;y_{1,j},y_{2,j}\in\Y_j$ where $y_{1,j}<y_{2,j}$ such that $\bar{f}(y_{1,j})=f(y_{1,j}),\;\bar{f}(y_{2,j})=f(y_{2,j})$. 
    Note that this condition holds when $\bar{f}$ is an SOS approximation of $f$. 
    Then, an upper bound on the error magnitude of the approximated composition over all of the regions
    \begin{align}
        \label{eqn:cor_unaryPWA_eq1}
        \max_{x\in\X}\ \abs{\bar{f}(\bar{g}(x))-f(g(x))}\ \leq \varepsilon_{f\circ g}
    \end{align}
    is given by
    \begin{align}
        \label{eqn:affine_err_prop}
        \varepsilon_{f\circ g}=\tau_f + \max_{y\in\underset{\tiny j}{\bigcup}\Y_j}\ \abs{f'(y)}\ \varepsilon_g\:.
    \end{align}
\end{cor}
\begin{pf}
    From Corollary~\ref{cor:PWA_err_bound}, it holds that
    \begin{align}
        \label{eqn:cor_unary_PWA_eq2}
        \max_{x\in\X}\ \abs{\bar{f}(\bar{g}(x))-f(g(x))}\ \leq \tau_f + d_{\bar{f}}\varepsilon_g.
    \end{align}
    By the mean value theorem, $\forall\;j=\{1,...,N\}\;\exists\;y^*_j\;$ where $y_{1,j},<y^*_j<y_{2,j}$, such that
    \begin{align}
        f'(y_j^*)=\frac{f(y_{2,j})-f(y_{1,j})}{y_{2,j}-y_{1,j}}=a_j\:.
    \end{align}
    Substituting into~\eqref{eqn:cor_unary_PWA_eq2} and using the fact that 
    \begin{align}
        \max_{j\in\{1,...,n_p\}}\ \abs{f'(y_j^*)}\ \leq\max_{y\in{\bigcup}_j\Y_j}\ \abs{f'(y)}
    \end{align}
    yields the desired result in~\eqref{eqn:cor_unaryPWA_eq1}.
    \qed
\end{pf}

Corollary~\ref{cor:MVT_err_bound_PWA} can be applied to yield an error bound for iteratively composed function approximations. For example, given the scalar function composition $f(g(h(x)))$, the approximation error bound is found to be
\begin{align}
    \varepsilon_{f\circ g\circ h} &= \tau_f+d_{f,g}\varepsilon_{g\circ h}\;, \\
    %%%
    &= \tau_f+d_{f,g}\big( \tau_g + d_{g,h}\varepsilon_h\big)\;, \\
    %%%
    &= \tau_f+d_{f,g}\tau_g + d_{f,g}d_{g,h}\big(\tau_h+d_{h,x}\varepsilon_x\big)\;, \\
    %%%
    \label{eqn:cor3_example}
    &= \tau_f+d_{f,g}\tau_g + d_{f,g}d_{g,h}\tau_h\;,
\end{align}
where $d_{f,g}=\max\abs{\frac{df}{dg}}$, $d_{g,h}=\max\abs{\frac{dg}{dh}}$, $\tau_f,\tau_g,\tau_h$ are the tolerances of $f$, $g$, and $h$, respectively, and $\varepsilon_x=0$ since here we assume that $x$ is known exactly. Note that~\eqref{eqn:cor3_example} is an affine function of the tolerances of each individual function, allowing the user to calculate the sensitivity of the composition error to each tolerance. 

\begin{exmp}
\label{ex:ErrProp_affOnly_2}
Again consider the function in~\eqref{ex-ErrPropAffOnly-eqn-y} over the domain $x\in[1,\ 3]$. The functional decomposition and bounds are given in \eqref{eqn:ex2_decomp}. The tolerances given in Table~\ref{tab:ExC1CompositionBounds_MOD} 
were used to construct PWA approximations for each function in the decomposition (via Method~1 in Section~\ref{sect:PWAApprox}). The error bounds for the composed approximation $\varepsilon_4$ are calculated using both Corollaries~\ref{cor:PWA_err_bound} and~\ref{cor:MVT_err_bound_PWA}, and are plotted alongside the composed PWA approximation in Figure~\ref{fig:cor2vscor3}. This example illustrates how Corollary~\ref{cor:MVT_err_bound_PWA} yields a more conservative (but quite similar) bound on the error, but only depends on the original functions and the unary approximation tolerances.

\begin{table}
    \centering
    \caption{Tolerances and propagated error bounds for the functional decomposition in Example~\ref{ex:ErrProp_affOnly_2}.}
    \begin{tabular}{c|c|c|c}
        Observable & $\tau_f$ & $\Bar{\varepsilon}_i$ & $\varepsilon_i$ \\ \hline\hline
        $w_2$ & $0.01$ & $0.01$ & $0.01$ \\ \hline
        $w_3$ & $0.01$ & $0.0186$ & $0.0194$ \\ \hline
        $w_4$ & $0.01$ & $0.0391$ & $0.0427$
    \end{tabular}
    \label{tab:ExC1CompositionBounds_MOD}
\end{table}

\newcommand\widV{2.75in}
\begin{figure}
    \centering
    \includegraphics[width=0.48\textwidth]{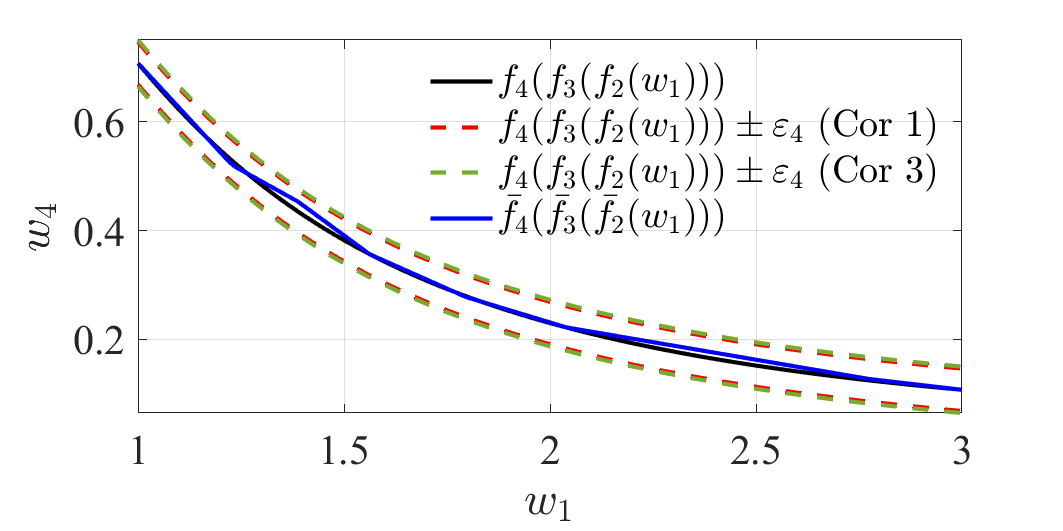}
    \caption{Using Corollary~\ref{cor:MVT_err_bound_PWA} yields a more conservative (less tight) error bound than Corollary~\ref{cor:PWA_err_bound}, but with the benefit of only depending on the original functions.}
    \label{fig:cor2vscor3}
\end{figure}
\end{exmp}

\section{Constructing an Approximation}
\label{sect:Examples}

The construction of a PWA approximation for a given function can generally be categorized into one of two problems. The first problem (which we call P1) is to minimize the number of breakpoints such that the maximum error satisfies a given tolerance. The second problem (which we call P2) is to minimize the error bound of the approximation given a maximum allowable number of breakpoints. 
Using the theory in Section~\ref{sect:ErrorProp}, and considering $f(g(h(x)))$ as~the function to approximate, these two problems can be written as:%
\begin{subequations}
\label{eqn:makeapprox_p1}
\begin{align}
    \text{P1}:\quad \min\ &n_f + n_g + n_h \\
    \st &\tau_f+d_{f,g}\tau_g + d_{f,g}d_{g,h}\tau_h \leq T\;, \label{P1constriant}
\end{align}
\end{subequations}
where $T$ is the tolerance to meet, and 
\begin{subequations}
\label{eqn:makeapprox_p2}
\begin{align}
    \text{P2}:\quad\min\ &\tau_f+d_{f,g}\tau_g + d_{f,g}d_{g,h}\tau_h \quad\quad\;\;\: \label{P2objective} \\ 
    \st & n_f + n_g + n_h\leq N\;,
\end{align}
\end{subequations}
where $N$ is the complexity limit.
As shown in in Figure~\ref{fig:NbreakJumpTol}, there is a direct relationship between $\tau_i$ and $n_i$ for each decomposed function, which can be leveraged in the solution of P1 and P2 because the result of Corollary~\ref{cor:MVT_err_bound_PWA} allows us to express constraint~\eqref{P1constriant} and objective~\eqref{P2objective} directly as a function of the tolerances. This is illustrated through the following numerical example.

\begin{exmp} 
\label{ex:OptTower} 
Consider the function
\begin{align}
    \label{eqn-y-source}
    y = f(x) = \sum_{i=1}^4 \frac{1}{d_i^2+1}\:,
\end{align}
where $f:[-5,\ 5]^2\rightarrow\Rspace,\; d_i=\norm{x-s_i}_2$ and 
\begin{equation}
    \label{eqn:tower_locations}
    s_1 = (1,3), s_2 = (-2,2), s_3 = (3,0), s_4 = (-1,-4)\:. 
\end{equation}% 
This function can be considered as the sum of multiple signals emitted from locations $s_i, \forall i\in\{1,2,3,4\}$, where each signal's strength decreases with the inverse square of the distance from the source. In~\cite{siefert2023_SVSE}, this function was considered in the context of set-valued state estimation. An SOS approximation was constructed via functional decomposition where each unary function was sampled over a uniformly spaced grid, and the resulting approximation had 163 breakpoints. The goal of this example is to construct a PWA approximation with the same complexity but a lower error tolerance. 
This corresponds to P2 above, where we leverage an error bound from Section~\ref{sect:ErrorProp} and Method 1 of Section~\ref{sect:PWAApprox} to relate tolerance and complexity. 

The decomposition of~\eqref{eqn-y-source} yields 28 unary functions, of which 4 are inverses of the form $\frac{1}{w+1}$, 8 are quadratics, and the remainder are affine, which can be represented exactly using PWA approximations. In this example, the CORA toolbox~\citep{CORA} was used to propagate the interval domains through each individual function. 

For each unary function, we logarithmically sample tolerances and find the number of breakpoints needed to meet those tolerances using Algorithm~\ref{alg:SOSstroll}, and substitute these values into the affine cost function. Then, we constrain the total number of breakpoints of the composition, which is just the sum of the individual numbers of breakpoints, to be at most 163. The resulting mixed-integer program is solved in 0.02 seconds to obtain a PWA approximation with 163 breakpoints and an error bound of 0.4453. Note that this is an upper bound on the approximation error, while further analysis places the true error at approximately 0.33 at maximum. The resulting approximation is thickened along the y-axis by the error bound and plotted in Figure~\ref{fig:TowerExample_Opt}(a). For comparison, Figure~\ref{fig:TowerExample_Opt}(b) shows the PWA approximation resulting from a uniform breakpoint sampling in \cite{siefert2023_SVSE}, which has a true error of approximately 0.60 at maximum. This illustrates that the proposed methods for intelligently placing breakpoints, guided by propagation of error through a functional composition, yield a significantly tighter fit to the true function as compared to a uniform spacing of breakpoints. 

\newcommand\widVI{.48\textwidth}
\begin{figure}
    \centering
    \begin{subfigure}[b]{\widVI}
        \centering
        \includegraphics[width=\textwidth]{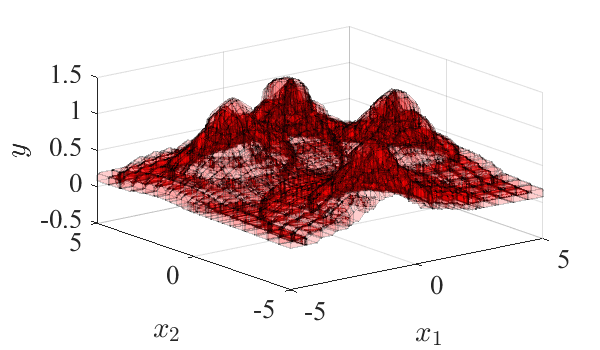}
        \caption{}
        \label{fig:TowerExample_Opta}
    \end{subfigure}\\
    \begin{subfigure}[b]{\widVI}
        \centering
        \includegraphics[width=\textwidth]{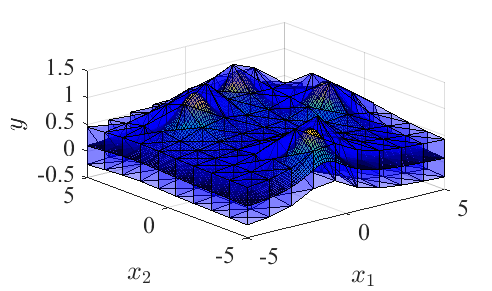}
        \caption{}
        \label{fig:TowerExample_Optb}
    \end{subfigure}\\
    \caption{PWA approximations of~\eqref{eqn-y-source}, both with 163 breakpoints. (a) Intelligent spacing of breakpoints via the proposed methods. (b) Uniform spacing of breakpoints.}
    \label{fig:TowerExample_Opt}
\end{figure}
\end{exmp}

\section{Conclusion}
\label{sect:conclusion}

This paper presents methods and algorithms to construct PWA approximations of a given complexity and/or error tolerance, and examines how those tolerances propagate through functional compositions. The theoretical results provide formal error bounds for composed PWA approximations of unary functions. Future work will apply these PWA approximation bounds, coupled with functional decomposition, to applications in neural networks, nonlinear reachability, and set-valued estimation. 

\bibliography{bibNew}
\end{document}